\vspace{.1in}\tolerance 200
\documentstyle[preprint,aps]{revtex}
\begin{document}
\draft

\title
{Enhanced Local Moment Formation in a Chiral Luttinger Liquid}
\author{Philip Phillips and Nancy Sandler }
\vspace{.05in}

%
\address
{Loomis Laboratory of Physics\\
University of Illinois at Urbana-Champaign\\
1100 W.Green St., Urbana, IL, 61801-3080}

%
\maketitle

\begin{abstract}
We derive here a stability condition for a local moment in the presence
of an interacting sea of conduction electrons.  The conduction electrons
are modeled as a Luttinger liquid in which chirality and spin are coupled. We show
that an Anderson-U defect in such an interacting system can be transformed
onto a nearly-Fermi liquid problem.  We find that correlations among the conduction
electrons stabilize the local moment phase.  A Schrieffer-Wolff transformation
is then performed which results in an anisotropic exchange interaction indicative
of the Kondo effect in a Luttinger liquid.  The ground-state properties of this model
are then equivalent to those of the Kondo model in a Luttinger liquid.    
\end{abstract}
\vspace{.1in}

\pacs{PACS numbers:72.10.Fk, 72.15.Nj, 75.20.Hr}

\narrowtext
Hybridization of a magnetic impurity with a band of conduction electrons results
in charge fluctuations that ultimately determine the stability of the local
moment on the defect.  If the defect is modeled with a single orbital that
can at most be doubly occupied, the Anderson-U model\cite{andersonU} 
\begin{eqnarray}
H_A & = & \sum_{k,\sigma}
\epsilon_{k} a_{k\sigma}^{\dagger} a_{k\sigma} + \sum_{\sigma} \epsilon_{d}
a_{d\sigma}^{\dagger} a_{d\sigma} + \sum_{k, \sigma} V_{kd} (a_{k\sigma}^{\dagger} a_{d\sigma}
+a_{d\sigma}^{\dagger} a_{k\sigma}) + U_d n_{d\uparrow} n_{d\downarrow}\\
& = & H_0 + \sum_{k,\sigma}
\epsilon_{k} a_{k\sigma}^{\dagger} a_{k\sigma}
\end{eqnarray}
is the simplest description
that includes charge fluctuations with the conduction electrons of the host 
metal.  In Eq. (1), $\epsilon_d$ is the defect energy of the magnetic impurity, 
$V_{kd}$ the
overlap integral between a band state with momentum k and the impurity, 
$a_{k}^{\dagger}$ creates an electron in the band states,
$a_{d\sigma}^{\dagger}$ creates an electron with spin $\sigma$ on the
impurity, and $n_{d\sigma} = a_{d\sigma}^{\dagger}a_{d\sigma}$ is the
number operator for an electron of spin $\sigma$.  As a consequence of the on-site 
repulsion ($U_d$), the single particle states on the impurity have energies, $\epsilon_d$ and
$\epsilon_d+U_d$. At high temperatures, the
density of states of this model has two Lorentzian peaks centered at these two energy
levels. However, the occupancy of these levels need not be equal.  Anderson\cite{andersonU} showed that at the Hartree Fock level,
$U_d \rho_d (0) = 1$ defines the phase boundary demarcating the magnetic (single occupancy) from the non-magnetic
(double occupancy) phase. In this relationship, $\rho_d$ is the defect 
density of states, $\epsilon_F = 0$ and we assumed that $\epsilon_d<0$. At low temperatures,
the local moment is quenched by the formation of 
the Kondo singlet state\cite{swolff}. Although the Kondo resonance is expected to occur for any value of the defect energy
within the range $-U_d<\epsilon_d<0$, it is most favourable at the defect energy corresponding
to the greatest stability of the local moment at the d-impurity, namely,
 $\epsilon_d=-\frac {U_d} {2}$. At this energy
$H_A$ is particle-hole symmetric, and the Kondo resonance is pinned at 
$\epsilon_F=0$.

If repulsive interactions are now turned on among the conduction
electrons, it is unclear at the outset whether they stabilize or destroy the
local moment phase of an Anderson-U impurity.  Should they destabilize the
local moment, then the possibility arises that the Kondo effect might be suppressed.
It is precisely the fate of a local moment in a sea of interacting
conduction electrons that we address in this paper.  Surprisingly, an answer to
this basic question has not
been advanced.  Magnetic impurities in heavy-Fermion and
high-T$_c$ materials are obvious physical realizations of an Anderson-U defect
in a strongly correlated system.  Hence,  formulation of the correlated local moment
problem is of extreme physical significance.  Recently, Schork and Fulde
\cite{schork} have considered
this problem in the limit that the defect is at most singly-occupied.  Consequently,
they were not able to obtain the phase diagram demarcating the magnetic from the
non-magnetic phase.  Others as well have considered the infinite-U limit or Kondo
problem in a correlated system\cite{fn}\cite{lt}\cite{si}.
The new physics that 
has come out of these studies is
that the Kondo temperature scales algebraically as $T_k\propto 
J^{2\pi v_F \over g_2}$, with J the Kondo exchange 
coupling,  $v_F$ the Fermi velocity and 
$g_2$ the coupling constant for forward
electron scattering.  This result is valid strictly in the limit of 
weak Coulomb interactions,
${g_2 \over 2\pi v_F }<<1$.  In addition, the Kondo effect occurs for both ferromagnetic
and antiferromagnetic exchange couplings.  In a recent paper, 
Schiller and Ingersent\cite{si}
have shown that if the conduction electrons are bosonized and only spin-up right movers
and spin-down left movers are retained, all weak coupling features of the Kondo problem
in a Luttinger liquid are reproduced, such as the power-law scaling of the 
Kondo temperature.  Below the Kondo temperature, the strong-coupling analysis of 
Furusaki and Nagaosa\cite{fn} results in anomalous exponents for the impurity
contribution to the specific heat.  The truncation of the full Luttinger liquid
made in ref. 6 results in purely Fermi-liquid behaviour.  Hence, the truncation
introduced by Schiller and Ingersent\cite{si} appears to be valid only in the weak
coupling regime.

In this paper, we approach the Anderson-U problem in a Luttinger liquid by coupling
chirality to spin as has been done previously in the work of 
Schiller and Ingersent\cite{si}.
We show here that with judiciously-chosen transformations, much of the physics of
this problem can be unearthed.  
Specifically, we show that interactions enhance the
local moment. As the local moment phase occurs in the weak coupling limit or
equivalently at temperatures high relative to the Kondo temperature, our conclusions
that the local moment phase is enhanced is most probably not affected by the 
truncation of the full Luttinger liquid.  We also show that in the Kondo limit, we recover exactly the
results of Schiller and Ingersent\cite{si}.

The starting point for our analysis is a Hamiltonian 
$H  =  H_0 +H_L$
that includes the 
standard Anderson model
as well as a 1-dimensional lattice of conduction electrons 
\begin{equation}\label{eq3}
H_L= -{t\over 2}\sum_{n\sigma} \left [ \Psi_{n\sigma}^{\dagger} \Psi_{n+1\sigma} +h.c.\right]
+U_L\sum_{n\sigma}\rho_{n\sigma}\rho_{n-\sigma}
\end{equation}
interacting via on-site Coulomb repulsions of strength $U_L$.  
In equation (\ref{eq3}), $t$ is the hopping matrix
element, $\Psi_{n\sigma}$ annihilates
an electron on site $n$ with spin $\sigma$ and $\rho_{n\sigma}=
\Psi_{n\sigma}^{\dagger} \Psi_{n\sigma}$ is the electron density at the n-th lattice site.
To obtain the Luttinger description of our interacting model, 
we now linearise around the Fermi momentum $k_F$ and write the electron
field as $\Psi_{n\sigma}=e^{ik_Fn}\Psi_{+\sigma}+e^{-ik_Fn}\Psi_{-\sigma}$.  
For each spin there are  two electron fields $\Psi_{\pm \sigma}$ with momentum $\pm k_F$.
We will, for the sake of simplicity, retain only two electron fields.
We associate with spin-up the right moving field, $\Psi_{+}$ and 
spin-down with the left-moving field, $\Psi_{-}$.  We refer to an interacting
system with this constraint as a chiral-spin Luttinger liquid.
This coupling of spin to 
chirality has been shown to have no severe consequences in the Kondo problem in a
Luttinger liquid\cite{si} in the weak coupling regime.   
Hence, it is worth exploring the Kondo limit
of the corresponding
Anderson model in our chiral-spin liquid. 
As there is only a forward interaction term, we can rewrite
our electron lattice Hamiltonian in continuum form as
\begin{equation}
H_L^{\prime}=-iv_F \sum_{s=\pm}s \int_{-L/2}^{L/2}\Psi^{\dagger}_{s}(x)\partial_x 
 \Psi_{s}(x)dx + aU_L\int_{-L/2}^{L/2} :\Psi_+^{\dagger}\Psi_+:
:\Psi_-^{\dagger}\Psi_-:dx
\end{equation}
where $::$ indicates normal ordering and $v_F$=atsin$k_F$a is the Fermi velocity and
$a$ the lattice spacing.  We must also
recast the impurity Hamiltonian
\begin{equation}
H_A^{\prime}=H_d +V_d\left[ \Psi_+^{\dagger}(x=0)a_{d\uparrow} + 
\Psi_-^{\dagger}(x=0)a_{d\downarrow} + h.c. \right].
\end{equation}
with $H_d$ the second and last terms in Eq. (1). In writing this
equation, we have subsumed the $\sqrt{a}$ continuum factor into
the definition of $V_d$.  Because spin and chirality are coupled, hopping of spin-up (spin-down) conduction
electrons to the impurity located at x=0 is mediated by the right (left) moving field
$\Psi_+(\Psi_-)$.

The lattice degrees of freedom can now be bosonized by writing the left and right moving
fields 
\begin{equation}
\Psi_{\pm}={1 \over \sqrt{2\pi a}}e^{-i\sqrt{\pi}
\left[\int_{-\infty}^x\Pi(x')dx'\mp\Phi(x)\right]}
\end{equation}
in terms of two real, conjugate Bose fields, $\Phi(x)$ and $\Pi(x)$.  The
Bose fields obey the commutation
relation $\left[\Phi(x),\Pi(y)\right]=i\delta(x-y)$. If we express the conduction
electron degrees of freedom in terms of the Bose fields, we find that the resultant
lattice Hamiltonian
\begin{equation}
H_L^{\prime}={v_F \over 2}\int_{-L/2}^{L/2}\left[ \alpha_-\Pi^2(x)+\alpha_+
(\partial_x\Phi(x))^2\right]dx
\end{equation}
can be mapped onto a free Fermion theory
\begin{equation}\label{eq5} 
H_L^{\prime}={\tilde{v}_F \over 2}\int_{-L/2}^{L/2}\left[ \tilde{\Pi}^2(x)+
(\partial_x\tilde{\Phi}(x))^2\right]dx
\end{equation}
once the bare Bose fields are rescaled in the form, $\tilde{\Pi}(x)=\eta^{1/2}\Pi(x)$
and $\tilde{\Phi}(x)=\eta^{-1/2}\Phi(x)$.
Such a rescaling retains the canonical
commutation relations. The constants appearing in the 
bosonized Hamiltonian are $\alpha_{\pm}=
\left[1\pm g\right]^{1/2}$,
$\alpha=\sqrt{\alpha_-\alpha_+}$, $\eta=\sqrt{\alpha_-/\alpha_+}$ and
a rescaled Fermi velocity $\tilde{v}_F=\alpha v_F$.  The strength
of the electron correlations is determined by the dimensionless conductance
$g={aU_L \over 2\pi v_F}$.  The range of validity of the approximations used here
is the weak-coupling regime, $g<1$.  In the rescaled basis,  interactions
among the conduction electrons simply rescale the original Fermi velocity.
It is this rescaling of the Fermi velocity that results in anomalous dimensions 
and as a consequence Luttinger liquid behaviour of the conduction electrons\cite{haldane}.

While the conduction electron degrees of freedom are now quite simple, the hybridization
term in the effective Anderson Hamiltonian
\begin{equation}
H_A^{\prime}  =  H_d+V_d\left[A_+^{\dagger}a_{d\uparrow}+ A_-^{\dagger}a_{d\downarrow} 
+h.c.\right]
\end{equation}
depends explicitly on the electron correlations through the operators
\begin{equation}
A_{\pm}^{\dagger}={1 \over \sqrt{2\pi a}}e^{i\sqrt{\pi}\left[\int_{-\infty}^0\eta^{-1/2}\tilde{
\Pi}(x')dx'\mp\eta^{1/2}\tilde{\Phi}(0)\right]}.
\end{equation}
We will have successfully mapped our impurity problem onto an equivalent non-interacting
one if the factors of $\eta$ in the exponents of $A_{\pm}$ can be rescaled to
unity.
We seek then a transformation that maps the fields $A_{\pm}$ to the canonical
rescaled fields
\begin{equation} 
\tilde{\Psi}_{\pm}^{\dagger}(x)={1 \over \sqrt{2\pi a}}e^{i\sqrt{\pi}\left[\int_{-\infty}^x\tilde{
\Pi}(x')dx'\mp\tilde{\Phi}(x)\right]}.
\end{equation}
>From the form of the hopping term in $H_A$, the transformation, should one exist,
must amount to a rotation among the states of the impurity.  To this
end, we investigate a unitary transformation of the form
\begin{eqnarray}
\hat{T} & = & e^{\left[i\Phi_+^{\lambda,\Gamma} n_{d\uparrow}+i\Phi_-^{\lambda,\Gamma} 
n_{d\downarrow}\right]}\\
\Phi_{\pm}^{\lambda,\Gamma} & = & \lambda^{-1}\int_{-\infty}^0\tilde
{\Pi}(x')dx'\mp\Gamma\tilde{\Phi}(0)
\end{eqnarray}
where
$\lambda$ and $\Gamma$ are constants to be determined. We need to evaluate 
$\hat{T}(H_L^{\prime}+H_A^{\prime})\hat{T}^{\dagger}=\tilde{H}$.  The terms in the Luttinger
part of the Hamiltonian transform straightforwardly as
\begin{eqnarray}
\hat{T}\tilde{\Pi}^2\hat{T}^{\dagger} & = &\left( \tilde{\Pi} +
\delta(x)\Gamma(n_{d\uparrow}-n_{d\downarrow})\right)^2\\
\hat{T}(\partial_x\tilde{\Phi}(x))^2\hat{T}^{\dagger} & = & \left( \partial_x
\tilde{\Phi}(x) + \delta(x)\lambda^{-1}(n_{d\uparrow}+n_{d\downarrow})\right)^2.
\end{eqnarray}
We will see that it is the $\delta(x)$ terms that  renormalize
the site energies as well as the impurity on-site Coulomb repulsion and produce a
Kondo exchange
interaction.  

The hopping term transforms as
\begin{equation}\label{eq6}
\hat{T}A_{\pm}^{\dagger}a_{d\pm\sigma}\hat{T}^{\dagger}= 
e^{-i\Phi_{\pm}^{\lambda,\Gamma}}A_{\pm}^{\dagger}a_{d\pm\sigma} + 
i[\Phi_{\pm}^{\lambda,\Gamma},A_{\mp}^{\dagger}]n_{d\pm\sigma}a_{d\mp\sigma}+ ...
\end{equation}
The second term in Eqn. \ref{eq6} represents a correlated hopping process
in which an electron is annihilated from a doubly-occupied impurity
state.  This term then
stabilizes the singly-occupied state of the impurity. The Hermitian conjugate of 
this term creates a doubly-occupied impurity state from a singly-occupied one.
The higher-order terms in Eq. \ref{eq6} are all of this form.  Hence, all higher-order
correlated hopping processes produce no net change in the occupancy of the impurity.
As a result, we drop these terms and focus solely on the first term in Eqn. \ref{eq6}.
The precise form of the transformation can now be found by demanding that
$e^{-i\Phi_{\pm}^{\lambda,\Gamma}}A_{\pm}^{\dagger}\propto
\tilde{\Psi}_{\pm}^{\dagger}(x)$.  Using the identity for combining products
of exponential operators, we find that
$\lambda^{-1}=\sqrt{\pi}(\eta^{-1/2}-1)$ and $\Gamma= \sqrt{\pi} (\eta^{1/2} -1)$.
We now combine the results of these calculations to obtain 
\begin{eqnarray}\label{eq7}
\tilde{H}&\approx&\tilde{H}_L^{\prime} + \tilde{\epsilon}_d(n_{d\uparrow}+n_{d\downarrow}) +\tilde{U}_d
n_{d\uparrow}n_{d\downarrow} +\tilde{V}_d\left[ \tilde{\Psi}_+^{\dagger}(x=0)a_{d\uparrow} + 
\tilde{\Psi}_-^{\dagger}(x=0)a_{d\downarrow} + h.c. \right]\nonumber\\
&+&\tilde{v}_F\Gamma
\tilde{\Pi}(0)(n_{d\uparrow}-n_{d\downarrow})+
\tilde{v}_F\lambda^{-1}\left(\partial_x\tilde{\Phi}(0)\right)(n_{d\uparrow}+
n_{d\downarrow})
\end{eqnarray}
as our fully-transformed Hamiltonian in the absence of the correlated-hopping
processes.  As is evident all bare parameters associated
with the impurity
have now been renormalized by the interactions:
\begin{eqnarray}\label{eq8}
\tilde{\epsilon}_d&=&\epsilon_d+{\tilde{v}_F\pi \over 2a}\left[(\eta^{-1/2}-1)^2+(\eta^{1/2} -1)^2\right]>\epsilon_d\nonumber\\
\tilde{U}_d &=& U_d+{\tilde{v}_F\pi\over a}\left[(\eta^{-1/2}-1)^2-(\eta^{1/2} -1)^2\right]>U_d\\
\tilde{V}_d&=&V_de^{i\pi/2(\eta^{-1/2}-\eta^{1/2})}\nonumber
\end{eqnarray}
Eqs. \ref{eq7} and \ref{eq8} represent one of
the key results of this paper.  Much of the physics of the Anderson model in 
a Luttinger liquid can be deduced from them.

Insight into the transformed Hamiltonian can be gained initially by ignoring the
last two terms in Eq. \ref{eq7}.  At this level, $\tilde{H}$ is identical to 
the original Anderson Hamiltonian except the matrix elements and site energies
now depend on the electron correlations through Eqs. \ref{eq8}.  
Note first that $\tilde{U}_d>U_d$. The interactions have effectively increased the on-site
Coulomb repulsions on the impurity, thereby increasing the energy cost of doubly
occupying the impurity.  The hybridization energy is unaffected by the interactions
because $|\tilde{V}_d|^2=|V_d|^2$.  As a consequence, the two crucial ratios that
determine the stability of the local moment phase
${\tilde{\epsilon}_d \over \tilde{U}_d}$ and 
${|\tilde{V}_d|^2 \over \tilde{U}_d}$ are both smaller than their bare values in
the Anderson model.  In the Anderson model, local moment formation is expected
for $0<|{\epsilon_d \over U_d}|<1$ and $\pi|V_d|^2/U_d<1$.  At this level of theory,
we find that the local moment phase must expand to compensate for the smaller values
of ${\tilde{\epsilon}_d \over \tilde{U}_d}$ and 
${|\tilde{V}_d|^2 \over \tilde{U}_d}$.  From this analysis, we conclude that
correlations stabilise the local moment phase.  

What effect do the last two terms
have?  Recall, $\tilde{\Pi}(0)=-\sqrt{\pi}j_1(0)$ defines the current and 
$\partial_x\tilde{\Phi}(0)=\sqrt{\pi}j_0(0)$ the electron charge density at the defect.  In. Eq.
\ref{eq7}, $j_1$ and $j_0$ couple respectively to the impurity magnetic moment
and charge density.  Above the Kondo temperature where local moment formation
is favoured, mean-field theory should be sufficient to describe the effect of 
these terms.  Physically, the charge density term should provide, to leading order,
an overall
shift in the defect site energy that scales with the filling. The net contribution of these terms to the local moment
phase is expected to be small as terms involving a product of two Fermion operators
is expected to yield corrections of $O(1/k_BT)$ at high temperatures\cite{lacroix}.
To see how this comes about, we calculate
the occupancy on the impurity level $\langle n_{d\sigma}\rangle$.
This quantity is obtained by integrating the imaginary part of the d-electron Green function,
$G_{d\sigma}(\omega)=\langle\langle a_{d\sigma}; a_{d\sigma}^{\dagger}\rangle\rangle$, weighted with
the Fermi-Dirac distribution function.  At high temperatures, it is sufficient
to use a second level closure of the equations of motion
\cite{lacroix} to eliminate
the conduction electron Green function $G_{kk^{\prime}}$ from the expression 
for $G_{d\sigma}$.  If we retain only the leading diagonal term in the equation
of motion for $G_{kk^{\prime}}$, we find that the occupancy on the impurity\cite{temp}
\begin{equation}\label{eq10}
\pi\langle n_{d\sigma}\rangle = cot^{-1}\left[ {\tilde{\epsilon}_{d}+
{\pi\tilde{v}_F\lambda^{-1}n\over a}-\tilde{\epsilon}_F+
\tilde{U}_{eff}^{\sigma}
\langle n_{d-\sigma}\rangle
-{\pi^2\Delta\tilde{v}_F \over 2LD}(\lambda^{-1}+sgn(\sigma)
\Gamma)\langle n_{d\sigma}\rangle
\over \Delta}\right]
\end{equation} 
is still of the mean-field Anderson form\cite{andersonU}. In this expression, the
energy cutoff is $D\propto k_BT$, $\Delta=\pi|\tilde{V}_d|^2\rho(\epsilon)/L$,
and $n=aN_e/L$ is the filling. 
We see explicitly now that the charge-density
term in Eq. \ref{eq7} provides a net shift to the defect site energy proportional
to the filling in the conduction.  Additionally, this term as well as the current
term provide corrections to the on-site Coulomb repulsion, 
$\tilde{U}_{eff}^{\sigma}=\tilde{U}_d- {\pi^2\Delta\tilde{v}_F \over 2LD}
(\lambda^{-1}-sgn(\sigma)
\Gamma)$ which are $O(1/k_BT)$. The overall effect of this correction is to 
decrease $\tilde{U}_d$.  However, it is straightforward to show that this correction
is always smaller than the enhancement in the on-site Coulomb repulsion predicted
by Eq.\ref{eq8}.
Hence, the enhnaced stability of the local moment remains intact even if the current
and density terms are included.  To illustrate the enhancement, we plot the phase
boundary in the limit that $D\rightarrow\infty$ in Eq. \ref{eq10}.  The Anderson
mean-field result $g=0$ corresponds to the solid line in Figure 1.
As is evident,
the local moment region expands as the strength ($g$) of the interactions among the 
conduction electrons increases.  As expected, the filling correction
provides a net shift in the defect site energy.  It appears then that the increased stability of the local moment phase can
be understood simply from the enhancement of the Coulomb repulsion induced by the 
electron correlations on the impurity. 

To understand the Kondo limit of our effective Hamiltonian, it is expedient
to introduce the anti-commuting Fourier components
$\tilde{\Psi}_{\pm}(x)={1\over \sqrt{L}}\sum_k e^{ikx}c_{k\pm\sigma}$. In terms of the
$c_k's$, the transformed Hamiltonian becomes,
\begin{eqnarray}\label{eq9}
\tilde{H}&=&\sum_k\epsilon(k)\chi_k^{\dagger}\sigma_z\chi_k +
\tilde{\epsilon}_d\chi_d^{\dagger}\chi_d+\tilde{U}_d
n_{d\uparrow}n_{d\downarrow}
+{\tilde{V}_d\over\sqrt{L}}\left[\chi_k^{\dagger}
\chi_d + h.c. \right]\nonumber\\ 
&+&aJ_z^U\sum_{k,k'} \left(\chi_k\sigma_z\chi_{k'}\right)\cdot\left(
\chi_d\sigma_z\chi_d\right)
+\sqrt{\pi}\tilde{v}_F\lambda^{-1} \sum_{k,k'}\chi_k^{\dagger}
\chi_{k^\prime}\chi^{\dagger}_d\chi_d
\end{eqnarray}
with $\epsilon(k)=kL\tilde{v}_F/2$,
$J^U_z=\tilde{v}_F\pi(1-\eta^{1/2})/a$, and $\chi_{k(d)}$ are 
two-components
spinors composed of the up and down components of $c_{k\sigma}(a_{d\sigma})$ 
scaled
by $1/\sqrt{L}$. 
It is the current density, $j_1(0)$, that gives rise to the new diagonal
spin-exchange process. This term
is identical to the one derived by Schiller and Ingersent\cite{si} in the context
of the Kondo effect in a chiral-spin Luttinger liquid.  
To see this more clearly, we perform a Schrieffer-Wolff transformation\cite{swolff}.  The hopping
term in Eq. \ref{eq9} produces the standard spin-flip term
$4|\tilde{V}_d|^2/\tilde{U}_d\left(\chi_k{\bf \sigma}\chi_{k'}\right)\cdot
\left(\chi_d{\bf \sigma}\chi_d\right)$.  As a result, the net diagonal spin-exchange matrix
element $\tilde{J}_z= 4|\tilde{V}_d|^2/\tilde{U}_d+ {\tilde{v}_F\pi\over a}(1-\eta^{1/2})$
consists of two contributions: the usual hybridization-mediated exchange
and an additional term arising from the interactions among the conduction electrons.
The perpendicular component of the exchange 
$J_\perp= 4|\tilde{V}_d|^2/\tilde{U}_d$ depends only on the hybridization-mediated interaction.  It is this anisotropy that is responsible for the algebraic
scaling of the Kondo temperature in a Luttinger liquid\cite{si}. 

Although an exact solution has not been found, 
it is reassuring that the chiral model is in agreement with simple 
physical arguments that electron correlations
enhance local moment formation. The success of the chiral-spin model is due
in part to the fact that it retains spin-flip backscattering at the impurity.
Lee and Toner\cite{lt} were first to point out 
that spin-flip backscattering is the dominant scattering process among 
the conduction electrons, at least for the Kondo problem.
Ultimately it would be preferrable to solve the local moment problem
in a full Luttinger liquid.  It turns out that a transformation that solves
the full problem is quite dis-similar from the one constructed here, 
because the bosonized Hamiltonian
for the full Luttinger liquid\cite{haldane} cannot be mapped onto a non-interacting Fermion
problem.  However, the charge and spin sectors independently satisfy
such a mapping, at least away from half-filling.  Hence, the analogous 
transformation must involve a rotation among the
charge and spin sectors of the Luttinger liquid as well as the states on the
impurity.

\acknowledgments
We thank Eduardo Fradkin, Yi Wan, Antonio Castro Neto, Myron Salamon, Abraham
Schiller and Kevin Ingersent 
for their characteristically level-headed remarks. 
This work is supported in part by the NSF grant No. DMR94-96134.

\centerline{\bf Figure Captions}

\bigskip

\noindent Figure 1: a) Local moment phase diagram calculated using Eq. \ref{eq10} with
$D=\infty$.  The curves indicate equal occupancy of the up and down spin
states on the impurity as a function of $x={\epsilon_d - \epsilon_f
\over U_d}$, $v=\Delta/U_d$, and the filling $n=aN_e/L$. The solid curve corresponds
to $g=0.0$ whereas $g=0.7$ in the other two. In all the curves $U_L=0.5U_d$.  
The local moment ceases to exist on and to the right of each curve.


\begin{thebibliography}{99 }
\bibitem{andersonU}
P. W. Anderson, Phys. Rev. {\bf 124}, 41 (1961).
\bibitem{swolff}
J. R. Schrieffer and P. A. Wolff, Phys. Rev. {\bf 149}, 491 (1966).
\bibitem{schork}
T. Schork and P. Fulde, Phys. Rev. B {\bf 50}, 1345 (1994).
\bibitem{lt} 
D. H. Lee and J. Toner, Phys. Rev. Lett. {\bf 69}, 3378 (1992).
\bibitem{fn}
 A. Furusaki and N. Nagaosa, Phys. Rev. Lett. {\bf 72}, 892 (1994).
\bibitem{si}
A. Schiller and K. Ingersent, Phys. Rev. B {\bf 51}, 8337 (1995).
\bibitem{haldane}
F. D. M. Haldane, J. Phys. C {\bf 14}, 2585 (1981); For a review see
J. Solyom, Adv. Phys. {\bf 28}, 209 (1979).
\bibitem{lacroix}
C. Lacroix, J. Phys. F {\bf 11}, 2389 (1981).
\bibitem{temp}
In deriving this expression, we assumed a flat Fermi distribution function.  As in
the mean-field analysis of ref. (1), our results are valid for temperatures
above the Kondo temperature but lower than those at which the second term in the 
Sommerfeld expansion becomes important.  For a typical metal, this is of $O(10^4K)$.
\bibitem{ludwig}
I. Affleck, Nucl. Phys. B {\bf 336}, 517 (1990).

\end{thebibliography}
\end{document}